\documentstyle[floats,aps]{revtex}
\begin{document}
\input epsf
\draft

\hsize=6.5truein
\hoffset=0.0truein
\vsize=9.0truein
\voffset=0.6truein
\hfuzz=0.1pt
\vfuzz=0.1pt
\parskip=\medskipamount
\overfullrule=0pt

\twocolumn[\hsize\textwidth\columnwidth\hsize\csname
@twocolumnfalse\endcsname
\title{Quantum Nucleation of Phase Slips in a $1d$ Model of a Superfluid}

\author{Jose A. Freire, Daniel P. Arovas and Herbert Levine}
\address{Physics Department, University of California, San Diego \\
La Jolla, CA 92093}

\date{\today}

\maketitle

\begin{abstract}
We use a $1d$ model of a superfluid based on the Gross-Pitaevskii Lagrangian
to illustrate a general numerical method designed to find quantum tunneling
rates in extended bosonic systems. Specifically, we study flow past
an obstacle and directly solve the imaginary time dynamics to find the
``bounce" solution connected with the decay of the metastable laminar state
via phase slip nucleation.  The action for the tunneling configuration goes
to zero at the threshold (in superfluid velocity) for classical production
of these slips. Applications to other processes are briefly discussed.
\end{abstract}

\pacs{PACS numbers: 64.60.Qb, 67.40.-w, 67.40.Vs}
\vskip2pc]

\narrowtext

The subject of quantum tunneling in extended systems continues to be of
great interest in a variety of condensed matter applications.  Examples
include superflows (either superfluids or superconductors) in which
dissipation can be induced by the creation of phase slips via tunneling.
Often, the rates of these processes are calculated by focusing on one relevant
degree of freedom (perhaps the center-point of a specific ansatz for the
quantum field) and approximately integrating out the remainder of the degrees
of freedom; this approach thereby reduces the problem to quantum mechanics of a
single particle.  This reduction precludes one from understanding the
extent to which re-arrangement of the (usually nonlinear) field degrees of
freedom can lead to more advantageous tunneling paths.

In this work, we offer a more direct approach to the evaluation of
quantum tunneling rates in extended bosonic systems. Specifically, we use
the idea popularized by Coleman and Callan that quantum tunneling can be
studied by finding a specific solution to the classical equations of motion
continued to imaginary time, the so-called ``bounce". To illustrate our method,
we explicitly study the nucleation of phase slips in a 1d model of a
superfluid flowing past an obstacle.  We show how the bounce solution can
be constructed, how it determines the nucleated state of the entire field
and how one can compute the action for this solution, thereby finding the
semiclassical rate.  At the end, we discuss extensions to other bosonic
systems as well as comment on the modifications that will be required to
study fermions.

To model a superflow with velocity $v$ past a $1d$ obstacle we use the
Gross-Pitaevskii Lagrangian density \cite{gross}
\begin{displaymath}
{\cal L} = i\hbar{\bar\psi} \partial_t \psi - {\hbar^2\over 2m}\,
|\partial_x \psi|^2 - {\textstyle{1\over 2}}g \left(|\psi|^2 - \mu(x)/g\right)^2
- i\hbar v {\bar\psi} \partial_x \psi\ .
\end{displaymath}
The first three terms arise directly from considering bosons
interacting with each other via $\delta$ function potentials of strength
$g$, in the presence of an obstacle treated as a position-dependent
chemical potential $\mu (x)$.   The last piece comes from working in a frame
in which the obstacle is at rest. We will explicitly choose $\mu(x)$ to have
the form of a square well 
of width $\Delta$ and depth $\eta \mu_0$ ($\eta < 1$), $\mu_0$
being the uniform chemical potential outside; this will give rise
to a depletion of the condensate density in the well.
In this model, the natural unit of length is the coherence length
$\xi=\hbar/\sqrt{m\mu_0}$, of velocity is the sound velocity, 
$c=\sqrt{\mu_0/m}$, and of condensate amplitude is $\sqrt{\rho_0}\equiv
\sqrt{\mu_0/g}$.  A dimensionless parameter,
$\alpha\equiv\rho_0\,\xi^d$ ($d=1$ here),
analogous to $1/\hbar$, then determines the importance of quantum
fluctuations; we assume $\alpha\gg 1$.
The classical dynamics derived from the above Lagrangian is the
Galilean-transformed non-linear Schr{\" o}dinger equation (NLSE).
In natural units, with $(\zeta(x)=\mu(x)/\mu_0)$,
\begin{displaymath}
i\partial_t\psi = - {\textstyle{1\over 2}}\partial^2_x \psi +
\left(|\psi|^2 - \zeta(x)\right)\psi -iv\partial_x\psi\ .
\end{displaymath}
In this equation, $\psi$ is often interpreted as representing ``the
wavefunction of the condensate''.  The spatial boundary condition is
taken to be one where the $\psi$ tends to a unimodular constant
far away from the obstacle.

In recent work by Frisch {\it et al \/} \cite{frisch}, the NLSE in
two spatial dimensions was utilized to study the problem of
$2d$ superflow past a circular obstacle. They found that there exists
a critical velocity such that for $v<v_{\rm c}$, one is able to find 
time-independent solutions corresponding to a laminar flow around the 
obstacle, whereas for $v>v_{\rm c}$ only time-dependent solutions exist.
The latter solutions correspond to a superflow that periodically nucleates 
vortex-antivortex pairs in the proximity of the disk; these pairs are
subsequently advected by the flow.  Similar results were found in
Ref. \cite{stone} for a superflow in a $2d$ channel containing a constriction.
The $1d$ analog of these vortex-antivortex pairs are phase slips.  These
are events localized in both space and time where the amplitude of the
condensate goes to zero and the phase slips by $2\pi$.  Accordingly, 
we investigated numerically the existence of time-independent solutions 
of this equation;  this was done by writing the time-independent
NLSE in a $1d$ grid and using Newton's method complemented by Powell's 
hybrid method \cite{powell,dennis} to solve the resulting system of 
nonlinear algebraic equations.  In direct analogy to the 2d findings,
we found that, associated with each well depth and width,
there is a critical value of the superflow velocity, $v_{\rm c}$,
above which such solutions do not exist.

The above result implies that the solution of the NLSE above $v_{\rm c}$ 
is time-dependent.  To find the nature of this solution, the time-dependent
NLSE was integrated using an operator splitting method \cite{numrec}
accurate to ${\cal O}(dt)$ and unconditionally stable.  An example of what
we found is shown in Fig. 1.  We observed the nucleation of the phase
slip, typically at the edge of the well where $|\partial_x\zeta|$ is
greatest, and the subsequent emission of a soliton.  This soliton, which 
propagates outside the well, is a member of a family of solutions of the 
NLSE with $\zeta=1$ \cite{soliton}.  Up to an overall phase,
\begin{displaymath}
\psi_{\rm u}(x,t) = \sqrt{1-u^2} \tanh \left[\sqrt{1-u^2}[x-(u+v)t]\right] +iu .
\end{displaymath}
This is a ``dark'' soliton: the amplitude dips to $|u|$ at its center.  
It carries a momentum $P=-2u\sqrt{1-u^2}$ and accounts for a phase slip of
$\pi-2\sin^{-1}u$ (see Fig. 1).

\begin{figure}[h]
\centering
\leavevmode
\epsfxsize=8cm
\epsfbox[18 144 592 718] {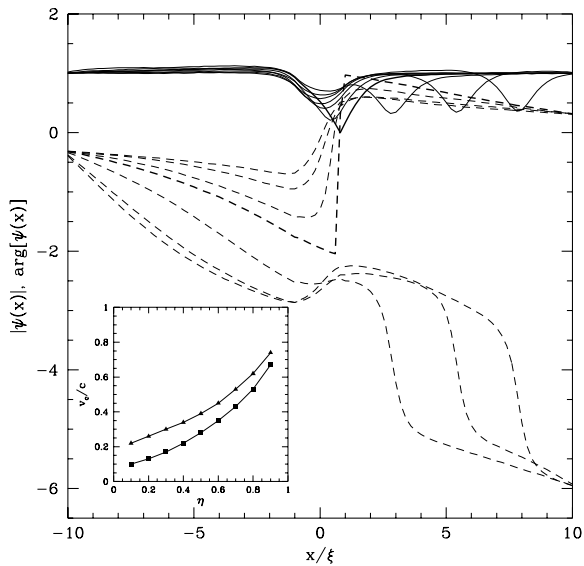}
\caption[]
{\label{fig:Fig1} The time evolution for $v>v_{\rm c}$, showing
amplitude (solid line) and phase (dashed line) of $\psi$ at 7 times between
$2\,\xi/c$ and $24\,\xi/c$. The phase slip occurs at $t\approx 12\,\xi/c$,
precisely at the right edge of the square well.  Afterwards, a soliton
is propagates to the right. The well width was $1.6\,\xi$ and the
chemical potential inside was $0.3\mu_0$.  The velocity was
$v=0.4\,c > v_{\rm c}=0.3\,c$.  $\psi$ was fixed at the ends.
Inset: critical velocity {\it versus\/} well depth for wells of width
$3.2\,\xi$ (squares) and $1.6\,\xi$ (triangles).}
\end{figure}

The existence of time-independent solutions of the NLSE only for $v<v_{\rm c}$
suggests a picture where the parameter $v$ creates a barrier that prevents the
{\it classical\/} decay of the metastable, time-independent state \cite{duan}.
This barrier disappears when $v>v_{\rm c}$ and thus only time-dependent
solutions can be found.  However, the steady flow state is only metastable
and therefore can decay via quantum fluctuations. This is similar to the
``fate of the false vacuum" problem studied by Coleman and Callan
\cite{coleman}. One can therefore use their formalism which relates the
the tunneling rate (to leading order in a semiclassical expansion) to the
classical action of the  periodic ``bounce solution", which proceeds in
imaginary time from the metastable state to the nucleated field and
back again. With the exception of one paper which studies nuclear fission
\cite{levit}, the possibility that one could directly solve for these bounce
solutions does not seem to have been previously realized.

In detail, we consider a coherent state path integral representation of
the matrix element of the evolution operator for imaginary time,
\begin{displaymath}
\langle \psi_0 | e^{-{\cal H}T/\hbar} | \psi_0 \rangle =\!\!\!\!\!\!
\int\limits_{\psi(-T/2)=\psi_0\atop{\bar\psi}(+T/2)=\psi_0^*}\!\!\!\!\!\!\!\!
{\cal D}[\psi,{\bar\psi}]\,\exp \left[-S_{\rm E}[\psi,
{\bar\psi}]/\hbar\right]\ .
\end{displaymath}
Here, $|\psi_0\rangle$ is the metastable coherent state wavefunction.
The Euclidean action is ($L$ is the spatial size of the system)
\begin{eqnarray}
S_{\rm E}[\psi,{\bar\psi}]&=&\rho_0\,\xi\int\limits_{-T/2}^{T/2}\!\!\! d\tau
\!\!\!\int\limits_{-L/2}^{L/2}\!\!\! dx\,
\left[ {\bar\psi}\partial_{\tau}\psi + E({\bar\psi},\psi) \right] , \\
E({\bar\psi},\psi)&=& {\textstyle{1\over 2}}
(\partial_x {\bar\psi})(\partial_x \psi)+ {\textstyle{1\over 2}}
[{\bar\psi}\psi - \zeta(x)]^2
- iv {\bar\psi} \partial_x \psi\ .
\end{eqnarray} 
The bounce solution is a classical configuration which extremizes this action;
we easily obtain the bounce equations for the fields $\psi$ and ${\bar\psi}$:
\begin{eqnarray}\label{eqn:bou}
-\partial_\tau \psi&=& -{\textstyle{1\over 2}} \partial^2_x \psi +
[{\bar\psi}\psi - \zeta(x)]\psi -iv\partial_x\psi \\ 
\partial_\tau {\bar\psi}&=& -{\textstyle{1\over 2}} \partial^2_x {\bar\psi} + 
[{\bar\psi}\psi - \zeta(x)]{\bar\psi} +iv\partial_x{\bar\psi}\ ,
\end{eqnarray}
with to $\psi(-T/2)=\psi_0$ and ${\bar\psi}(T/2)=\psi_0^*$.
The spatial boundary conditions are the same as before. These equations imply
${\bar\psi}(\tau)=\psi(-\tau)^*$, hence ${\bar\psi}(\tau)=\psi^*(\tau)$ -- the
density ${\bar\psi}\psi$ is real --  only at three time slices:
$\tau=\pm{\textstyle{1\over 2}}T$, when $\psi=\psi_0$, and $\tau=0$, when
the field emerges from the barrier and the phase slip has been nucleated
\cite{jain}.  It is crucial to recognize that we do not need to make any
assumption regarding the nature of the nucleated state, as the solutions of
the equations will directly and uniquely determine this field configuration.

These equations can not be solved by integrating forward in time. This is
because the ${\bar\psi}$ field has negative diffusivity and hence forward in
time integration is ill-posed; alternatively, eliminating ${\bar\psi}$ using
the above relation will give an explicitly non-local (in time) system.
Even were this not to be the case, the fact that the bounce is a saddle-point
as opposed to a minimum renders time-integration difficult. Instead, we
solved those equations iteratively by discretizing them on a {\em space-time}
lattice and using the Newton-Powell method mentioned above to find the
zero of the resulting system of nonlinear algebraic equations. This iterative
scheme does not care about time-locality, nor does it assume stability. 

\begin{figure}[t]
\centering
\leavevmode
\epsfxsize=8cm
\epsfbox[18 144 592 718] {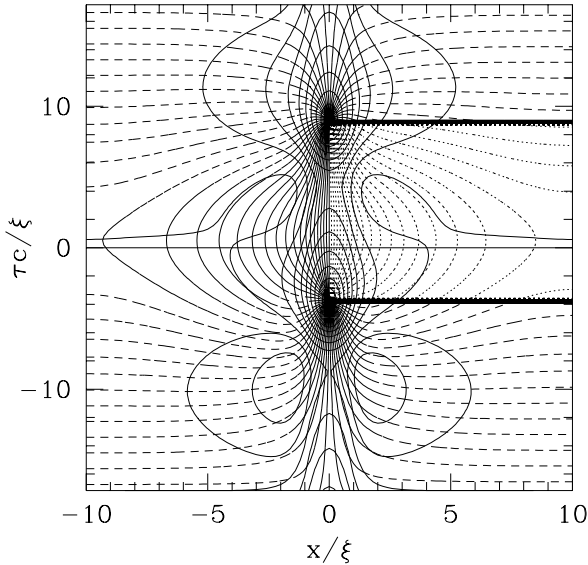}
\caption[]
{\label{fig:Fig2} Contour plot of the amplitude (solid) and phase (dashed)
of $\psi(x,\tau)$ for the bounce solution.  Here $v=0.1\,c$ while
$v_{\rm c}=0.3\,c$. Heavy lines are where the phase crosses $-2\pi$ and do
not correspond to any singularity.  The well parameters are the same as in
Fig. 1.}
\end{figure}

In solving the bounce equations on a space-time lattice one must choose a
time interval $T$ long enough to include all the structure around
$\tau=0$, as it is that part of the bounce that contributes the most to
the action and that gives the nucleated state.  We set up the equations
in such a way that at $\tau=-T/2$ we had the exact metastable state, while
at once we also enforced the relation ${\bar\psi}(\tau)=\psi^*(-\tau)$.
Because $T$ is finite, the solution obtained is not exactly periodic -- $\psi$
at $\tau=T/2$ is slightly different from $\psi$ at $\tau=-T/2$ -- but the
difference can be made arbitrarily small by increasing $T$. The energy of the
solution, however, was nearly constant and equal to the metastable state
energy, as is the case for the true bounce solution.
We also work with a finite spatial size.  Whereas in the infinite 
system size limit we can have both $\psi \rightarrow e^{i\theta_0}$ and
$\partial_x \psi=0$ as $x \rightarrow \pm \infty$, in our case we had
to choose between these two options.  We found that for smaller system sizes
the gradient condition worked best.  Typically we had $T \sim 34\, \xi/c$ and
$L \sim 20\, \xi$ in a grid of 100 spatial points by 50 time points. This
resulted in a system of 10,000 real nonlinear algebraic equations for
the real and imaginary parts of $\psi$.  The iterative scheme converges
rapidly to a well-behaved solution when an initial guess with the
correct topology is used to start the process.

As expected, we found a solution that precisely corresponds to the quantum
nucleation of a phase slip from the time-independent state of the given
well at superflow velocity ($v < v_{\rm c}$).  The field $\psi(x,\tau)$
corresponding to a phase slipped state at $\tau=0$
has a singular point in negative time around which its phase winds
counter-clockwise and another at positive time with clockwise phase winding.
This vortex-antivortex pair in space-time results in a state at
$\tau=0$ with a phase slip of $2\pi$ with respect to the metastable state at
$\tau=\pm T/2$.  The amplitude and phase of a typical solution $\psi$ are
shown in Figs. 2 and 3.  Note that the phase singularities are not
symmetrically displaced about $\tau=0$ -- $\psi$ and ${\bar\psi}$ are only
related by complex conjugation for $\tau=0,\pm T/2$.
Precisely at $\tau=0$ the configuration is
interpretable as one for the ordinary ({\it i.e.\/} real-time) NLSE and hence
can be treated as the initial condition for the subsequent real-time
evolution of the nucleated phase slip.  When this is done, the result
is quite similar to what we have already seen occurs for the classically
nucleated slip at $v > v_{\rm c}$; that is, a soliton is formed
which propagates to the right, transferring energy from the
obstacle region towards infinity.

\begin{figure}[t]
\centering
\leavevmode
\epsfxsize=9cm
\epsfbox[18 144 592 718] {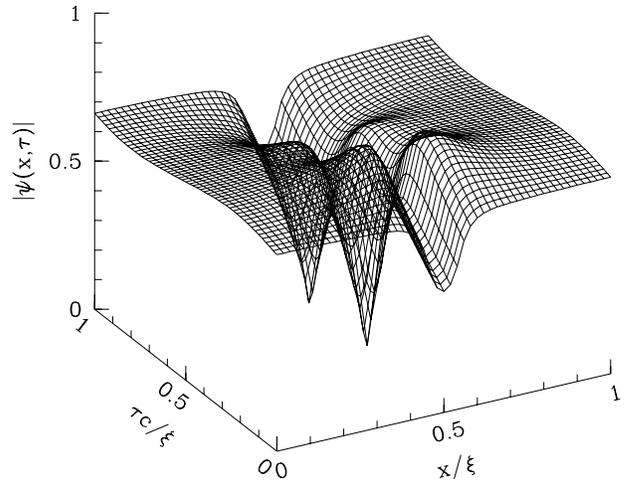}
\caption[]
{\label{fig:Fig3} The surface plot of the amplitude of the field $\psi$
of Fig. 2.  The depletion of the condensate inside the well and in the vortex
cores is apparent.}
\end{figure}

The decay rate of the metastable state can be found using $\Gamma \propto
\exp [-\Delta S_{\rm E}/\hbar]$, where $\Delta S_{\rm E}$ is the difference
between the bounce action and the metastable state action. Since the
dynamics in imaginary time conserves energy, we have
\begin{equation}\label{eqn:gam}
\Gamma \propto \exp \left( -\rho_0\,\xi\int\!\! d\tau\!\!\int\!\! dx\,
{\bar\psi} \partial_\tau \psi \right) .
\end{equation}
We solved the bounce equations for various superflow velocities with a
given well and computed $\Gamma$ according to equation \ref{eqn:gam}.
The behavior is depicted in Fig. 4; note that the action
approaches zero as we reach the classical threshold.  As $v$ approaches
$v_{\rm c}$ the vortex-antivortex pair separation decreases until eventually no
time-dependent solution of equations \ref{eqn:bou} 
can be found for $v>v_{\rm c}$. 

\begin{figure}[h]
\centering
\leavevmode
\epsfxsize=8cm
\epsfbox[18 144 592 718] {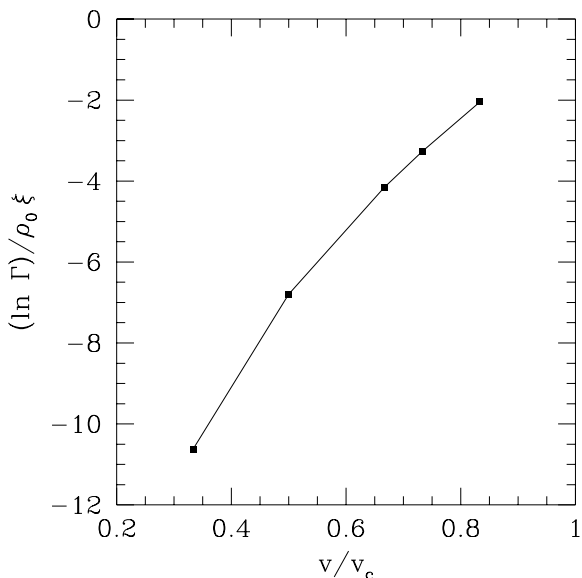}
\caption[]
{\label{fig:Fig4} An example of the dependence of the semiclassical decay rate
on the superflow velocity.  The well is the same as Figs. 2 and 3.
For $v>v_{\rm c}$ no metastable state exists.}
\end{figure}

Although we have described our method for a specific example, it should be
clear that the same approach can be used to compute tunneling rates and
nucleated states for any bosonic system.  Applications we have in mind
include higher dimensional superfluid-obstacle problems (such as the quantum
analog of the 2d classical nucleation of vortex-antivortex pairs discussed
above) and the decay of the metastable Bose-Einstein condensate state for
trapped alkali atoms with {\it attractive\/} interactions \cite{BEC}.  One can
also imagine straightforward extensions to the traditional field of macroscopic
quantum tunneling where one could study the tunneling of, say, a two level
system coupled to a {\it nonlinear\/} set of continuous degrees of freedom.
Finally, extensions to fermionic systems (either charged or uncharged) are more
challenging.  Integrating out the fermions typically leaves one with a
bosonic action in the form of some functional determinant and the resultant
equation of motion involves successfully evaluating the inverse of a complicated
operator. Whether it is feasible to directly find the bounce solution for
this case (with no further approximations) is currently under investigation.

We wish to thank J. Koplik and S. Renn for valuable discussions.
HL is supported in part by NSF Grant DMR94-15460.

\end{document}